\documentclass[%
 aip,
 amsmath,amssymb,
preprint,%
]{revtex4-1}

\usepackage{graphicx}
\usepackage{dcolumn}
\usepackage{bm}

\usepackage[utf8]{inputenc}
\usepackage[T1]{fontenc}
\usepackage{mathptmx}
\usepackage{etoolbox}

\makeatletter
\def\@email#1#2{%
 \endgroup
 \patchcmd{\titleblock@produce}
  {\frontmatter@RRAPformat}
  {\frontmatter@RRAPformat{\produce@RRAP{*#1\href{mailto:#2}{#2}}}\frontmatter@RRAPformat}
  {}{}
}%
\makeatother
\begin{document}

\preprint{AIP/123-QED}

\title{Neural ODE to model and prognose thermoacoustic instability}
\author{Jayesh M. Dhadphale}
\email{jayeshmdhadphale@gmail.com}
\affiliation{Department of Aerospace Engineering, Indian Institute of Technology Madras,
Chennai, Tamil Nadu 600036, India}
\author{Vishnu R. Unni}%
\affiliation{
 Department of Mechanical and Aerospace Engineering, University of California San Diego, La Jolla, CA 92093, United States 
}%
\author{Abhishek Saha} 
\affiliation{
 Department of Mechanical and Aerospace Engineering, University of California San Diego, La Jolla, CA 92093, United States 
}%
\author{R. I. Sujith}
\email{sujith@iitm.ac.in}
\affiliation{Department of Aerospace Engineering, Indian Institute of Technology Madras,
Chennai, Tamil Nadu 600036, India}

\date{\today}

\begin{abstract}
Thermoacoustic instability in reacting flow field is characterized by high amplitude pressure fluctuations driven by a positive coupling between the unsteady heat release rate and the acoustic field of the combustor. For turbulent flow, the transition of a thermoacoustic system from a state of chaos to periodic oscillations occurs via a state of intermittency. During the transition to periodic oscillations, the unsteady heat release rate synchronizes with the acoustic pressure fluctuations. Thermoacoustic systems are traditionally modeled by coupling the model for the heat source and the acoustic subsystem, each estimated independently. The response of the unsteady heat source, the flame, to acoustic fluctuations is characterized by introducing unsteady external forcing. The forcing necessitates a powerful excitation module to obtain the nonlinear response of the flame to acoustic perturbations. Instead of characterizing individual subsystems, we introduce a neural ordinary differential equation (neural ODE) framework to model the thermoacoustic system as a whole. The neural ODE model for the thermoacoustic system uses time series of the heat release rate and the pressure fluctuations, measured simultaneously without introducing any external perturbations, to model their coupled interaction. Further, we use the parameters of neural ODE to define an anomaly measure that represents the proximity of system dynamics to limit cycle oscillations and thus provide an early warning signal for the onset of thermoacoustic instability.
\end{abstract}

\maketitle

\begin{quotation}
Energy conversion devices such as gas turbine engines, rocket engines have combustors as an integral part which confines the reacting flow field. The acoustic field inside the combustor interacts with the reacting flow field inside the combustor. The positive coupling between acoustics and reacting flow field can result in high amplitude pressure oscillations known as thermoacoustic instability. Thermoacoustic instabilities have catastrophic consequences such as premature failure of engine components, leading to costly shut down of land-based gas turbine engines used for electricity generation. For rocket engines, thermoacoustic instability can overwhelm the thermal protection system, affect onboard electronics and guidance systems, and endanger the mission's success. Thus, the prediction of onset and amplitude of thermoacoustic instability is an intriguing problem for gas turbine and rocket engines. Reduced ordered models are often used to study thermoacoustic systems where the model for individual subsystems such as acoustic and flame response are obtained independently. The requirement of a powerful excitation system to obtain a nonlinear response of the flame to velocity fluctuations limits the application such reduced ordered model to test rigs. The number of state variables that can be measured is often limited due to the inaccessibility of the combustor in situ, which makes it difficult to access the high dimensional dynamics of the system. We address these issues by combining a method of delay embedding with neural ODE, to model the thermoacoustic system. We model the thermoacoustic system as a whole by training the neural ODE to learn the coupled interaction of unsteady heat release rate and pressure fluctuations from their simultaneously measured time series. Further, using the parameters of neural ODE, we also propose a precursor measure to predict the onset of the oscillatory instability in the thermoacoustic system.
\end{quotation}

\section{Introduction}
\label{sec:introduction}
Nature abounds with systems that exhibit oscillatory instabilities. Such instabilities are found in many fluid dynamic systems where we observe the emergence of self-sustained oscillations upon variation of some control parameters of the system. Examples include, thermoacoustic \citep{Juniper_Sujith-2018}, aeroacoustic \citep{Flandro_2003} and aeroelastic \citep{Hansen2007} instabilities. These oscillatory instabilities, characterized by spatially and temporally ordered behavior, emerge as a result of positive coupling between fluctuations in different subsystems that comprise the system. Such coupled interactions are often nonlinear in nature. Further, many practical systems that engender oscillatory instabilities also have an associated turbulent fluid flow, which adds further complexity to the interaction between different subsystems. Turbulent flows comprise eddies characterized by a wide range of length and time scales. There is the transfer of energy across these scales occurring through various cascade mechanisms \citep{Lewis1926, Kraichnan-1967}. These cascade processes introduce perturbations in the coupled interactions between subsystems of a fluid dynamic system. Since there are interactions of different scales and nature in such turbulent fluid dynamic systems, one can regard such a system as being composed of multiple interacting subsystems. A system comprising multiple interacting subsystems of wide-ranging scales and exhibiting unique collective behavior as a result of this interaction is known as a complex system~\citep{Smith2003ComplexSI}. The objective of this paper is to model the interaction of subsystems of a complex system such as a turbulent fluid dynamic system responsible for the onset of oscillatory instability using neural ordinary differential equations. To elaborate on this modeling approach, in this paper, we discuss the estimation of the neural ODE model of a thermoacoustic system exhibiting oscillatory instability. 

In thermoacoustic systems, the ordered oscillatory behavior that emerges due to nonlinear coupling between the heat source and acoustic field can result in high amplitude pressure and heat release rate oscillations \citep{Juniper_Sujith-2018}. Enhanced heat transfer due to thermoacoustic instability can overwhelm the thermal protection systems in rocket engines and introduce thermal stresses in combustor walls \cite{Lieuwen-2005}. Cyclic loading due to high amplitude pressure oscillations can result in premature fatigue failure of the structural components in gas turbine and rocket engines. Such premature failure can result in a costly shutdown of the gas turbine engines that are used for electricity generation \cite{Lieuwen-2005}. For rocket engines, severe vibrations induced by thermoacoustic instability could affect the onboard electronics, and the navigation and control of the rocket, eventually leading to mission failure \cite{Lieuwen-2005, Rem-Giants}. Considering the catastrophic consequences of high amplitude oscillatory instability, real-time monitoring and estimating the proximity of the system to the onset of thermoacoustic instability is of paramount importance. 

 Detection of the onset of thermoacoustic instability needs measures that can quantify the characteristic changes in the system during the transition from a stable state of a combustor characterized by low amplitude broadband noise (combustion noise) to a high amplitude oscillatory state. This transition is associated with an emergence of oscillatory motion with characteristic frequencies, accompanied by an increase in the amplitude of oscillation of state variables (acoustic pressure and acoustic velocity). The rising amplitude of oscillations can be detected using the root mean square (RMS) measure whereas the peak in the amplitude spectrum corresponding to the characteristic frequencies can be obtained using the fast Fourier transform (FFT) of the time series of the state variable.

Considerable efforts have been expended in developing precursors based on an understanding of the dynamic nature of the thermoacoustic systems before the onset of thermoacoustic instability  \citep{SUJITH2020}. The stable operation of turbulent combustion systems generates aperiodic acoustic pressure oscillations, which are traditionally known in the thermoacoustic parlance as combustion noise. Using statistical tests for determinism, \citet{Vineeth_Loss_of_chaos} and \citet{Tony-2015} showed that combustion noise has a deterministic signature and that the thermoacoustic system is in a state of high dimensional chaos during the occurrence of combustion noise. During this state, time series of acoustic pressure displays multifractal characteristics, suggesting the presence of multiple interacting time scales in the system. However, the emergence of order in the system is associated with the loss of this multifractal nature and the emergence of a dominant temporal and spatial scale. Thus multifractal spectrum width acts as an early warning measure for the onset of oscillatory instability \citep{nair_sujith_2014, VENKATRAMANI2017390}. The loss of multifractality also implies that the transition of the thermoacoustic system from chaos to ordered limit cycle oscillations features a reduction in the system complexity. The complexity of the time series of pressure fluctuations can be quantified using permutation entropy. \citet{Gotoda_Precursor_Per_Entropy_2017} applied a modified permutation entropy to detect the onset of a shift in the dominant frequency mode before the amplification of pressure fluctuations.

Time series obtained from dynamical systems can also be analyzed using complex networks. A complex network represents the dynamical system as a network of nodes. Nodes are connected by links and form a network following a predefined rule. Visibility graphs \cite{Lucas_Visibility_2008}, a tool from network theory, can be used to convert time series into a network representing the patterns in the time series. The constructed visibility graph reflects several properties of the time series. Regular, random, and scale-free graphs respectively reflect the periodic, random and fractal nature of the time series. \citet{Meenatchidevi_2015} detected the onset of thermoacoustic instability as a transition of the topology of a reconstructed visibility graph from scale-free to a regular graph. \citet{Gotoda_Precursor_Motif_2018} combined the framework of computing motif profiles associated with horizontal visibility graph and principal component analysis to detect the onset of thermoacoustic instability.

The occurrence of intermittency during the transition from chaotic to periodic state is the key feature in the emergence of oscillatory instabilities in turbulent systems \citep{nair_thampi_sujith_2014, Nair-2016, Venkatramani-2016}. Intermittency features epochs of ordered oscillations amidst
a periodic oscillations. \citet{nair_thampi_sujith_2014} used various recurrence measures such as the recurrence rate, Shannon entropy, and trapping time to detect the onset of oscillatory instability. A study of synchronization between the acoustic pressure and the heat release rate fluctuations shows that these signals intermittently synchronize during the occurrence of intermittency \citep{pawar_seshadri_unni_sujith_2017}. During the occurrence of intermittency, ordered and disordered structures coexist in the spatial domain, resembling a chimera state \cite{mondal_unni_sujith_2017}.

The advances in experimental techniques and high-performance computing have made an abundance of data available and finding patterns out of huge data is a challenging task. A systematic framework is necessary to extract useful information from data. Machine learning provides a framework to systematically detect patterns in the data. Pattern recognition techniques such as dynamic mode decomposition (DMD), proper orthogonal decomposition (POD), and deep neural networks can be applied to detect the formation of ordered structures in the flow \citep{Brunton_ann_rev_2020}. The operating state of a thermoacoustic system can be characterized by symbolic time series analysis.  In symbolic time series analysis, first, the time series of a state variable is appropriately converted into a symbolic series; consequently, the patterns in the time series are analyzed by studying the statistics of words (strings of symbols) in the symbolic series. Different approaches to construct symbolic time series from experimental data are presented in  \citet{Symbolic_timeseries_review_Daw_2003}. \citet{Sarkar_Ray_2010}, \citet{vishnu_symbolic_2015} and \citet{Sarkar_2015}  applied symbolic time-series analysis to detect the onset of transition to oscillatory instability with probabilistic measures.

The stability of combustors can be assessed experimentally or by numerical simulation. Iterative optimization using experimental and numerical methods is very expensive and numerical solutions are very slow for real-time control \citep{Bruton_2015}. Reduced-order models capture the essential flow dynamics at a fraction of the cost and considerable efforts have been expended towards developing them \citep{Rowly_ann_review_2017}. Machine learning provides a framework to systematically incorporate data available from experiments and numerical simulations to reduce the dimensionality of the data and construct reduced-order models.

The identification of a dynamical system from data involves finding a function that captures the essential dynamics of the system. To approximate this function, we use neural networks. A computational system that imitates the behavior of the network of neurons in a biological brain is known as a neural network (NN). According to the universal approximation theorem \cite{Kur_1989_Universal_Approx}, any function can be approximated by a sufficiently deep NN which makes a NN the natural choice for modeling dynamical systems.

The main objective of the present study is to model the coupled behavior of $p'$ and $\dot{q}$, obtained without imposing any external acoustic fluctuations and develop a precursor measure to quantify the proximity of the state of the system to the onset of oscillatory instability. Recently, \citet{Chen_NODE_2018_7892} have proposed a methodology to model differential equations of time series data using NN termed as the neural ordinary differential equations (neural ODE). We train the neural ODE on the experimental data obtained from a thermoacoustic system; i.e., time series of pressure and heat release rate oscillations, to obtain a neural ODE model to approximate the underlying dynamical system. Here, we seek the interdependence of state variables for the thermoacoustic system such as heat release rate and pressure fluctuations using neural ODEs. Resorting to data-driven model identification obviates the need for explicit modeling of the driving function, boundary conditions and damping effects. Training the neural ODE, we obtain the parameter vector $\theta$ for each operating state decided by the control parameter such as the Reynolds number ($Re$) or the equivalence ratio. We analyze the evolution of the thermoacoustic system in the parametric space where the elements of the vector $\theta$ represent the coordinates of the system. Here, we are indeed quantifying the proximity of the system to the onset of limit cycle oscillations in the parametric space and propose a precursor measure to detect the onset of oscillatory instability.

The paper is organized as follows. A brief overview of NN and neural ODE along with thermoacoustic system modeling and the precursor measure are presented in Sec. \ref{sec:model_description}. In  Sec. \ref{sec:experiments}, we describe experiments performed to get data for modeling the thermoacoustic system and to validate the proposed precursor measure. Results obtained from the current framework are discussed in Sec. \ref{sec:results_discussion}. The key findings are then summarized in Sec. {\ref{sec:conclusion}}. The generalizability of the current framework to model systems that show the transition from chaos to periodic oscillations is demonstrated for an aeroacoustic system in Appendix \ref{res_generlizability}.  

\section{Description of the neural network framework}\label{sec:model_description}

\begin{figure}[ht!]
    \centering
     \includegraphics[width=0.9\textwidth]{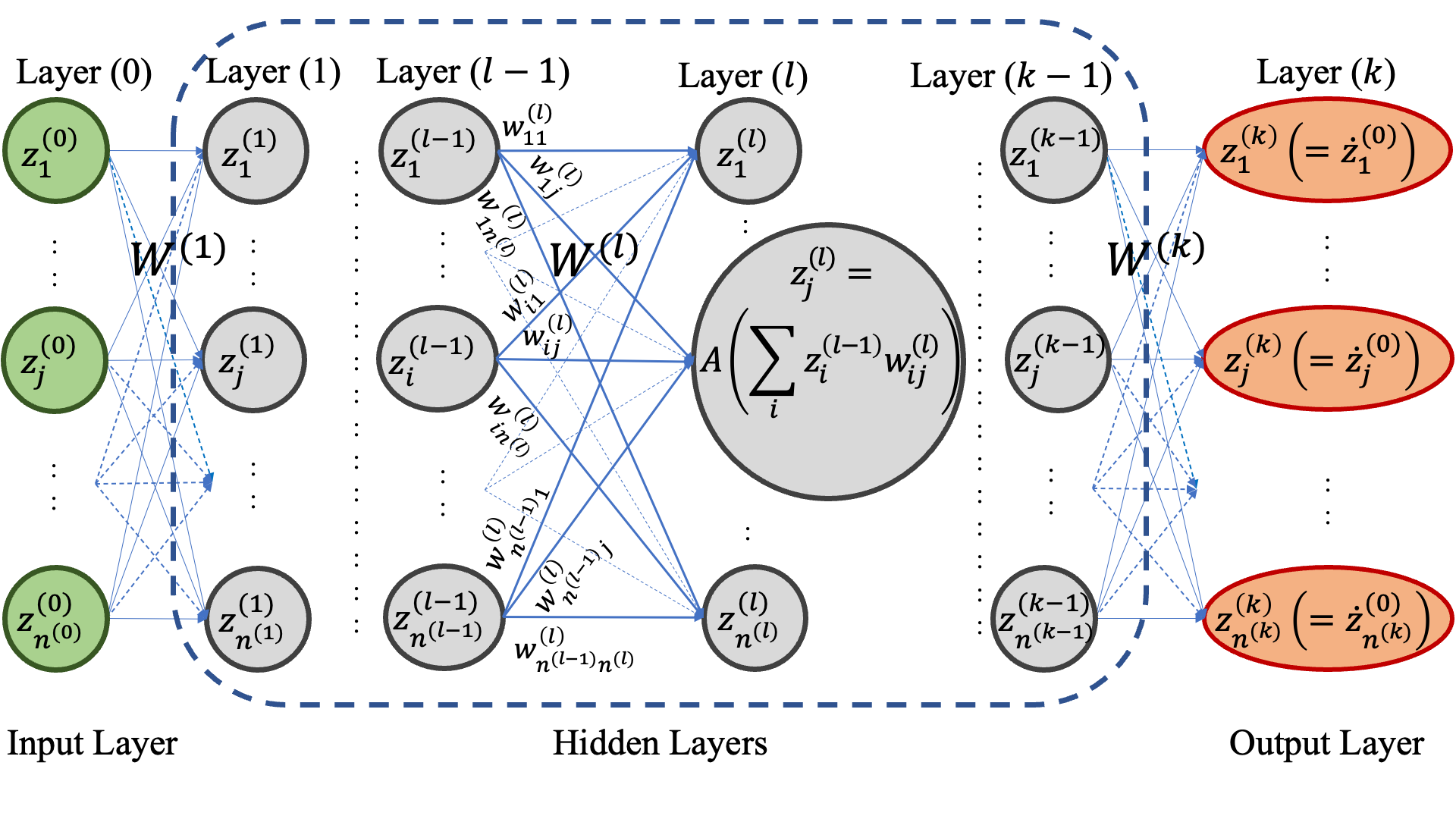}
     \caption{Schematic of NN with $(k-1)$ hidden layers. Layer (0) is the input and Layer $(k)$ is the output layer. The $l^{th}$ layer contains $n_{(l)}$ number of nodes. The $z^{(l)}_j$ is the output of $j^{th}$ node in $l^{th}$ layer.  $w^{(l)}_{ij}$ is the weight assigned to the link that connects $i^{th}$node from layer $(l-1)$ to $j^{th}$ node in layer $(l)$.  $W^{(l)}$ is the weight matrix that maps the output of layer $(l-1)$ to the input of layer $(l)$ and $A$ is the activation function which maps the linear combination of inputs to the output of node.  For neural ODE the nodes in the output layer gives the time derivative of the corresponding input node variable; i.e.,  $z^{(k)}_j=\dot{z}^{(0)}_j$.}
     \label{fig:nn_ode}
\end{figure}
Artificial NNs are computational systems that approximately mimic the behavior of a network of neurons in an animal brain. They are constituted by connected units known as artificial neurons or nodes. A node can transmit a signal across a connection to another node in a manner similar to how information flows through synapses of a biological brain. This signal is further processed by a second node and is then transmitted to other nodes connected to it. The `signal' transferred between nodes is a real number and the output of a node is a  nonlinear mapping of a linear combination of all the inputs to a node. The connections between the nodes are called edges. These edges have weights that are adjusted as the NN undergoes the process of learning, otherwise known as training. The weight of an edge modifies the strength of the signal from one node to another. The neurons are often arranged in layers, each of which transforms the input information in a different manner. The input information thus passes through a sequence of layers, to obtain a desired output according to the objective of the training. The objective of the training could be to classify the input data, to identify patterns in the data, make predictions based on the input data and so on. In the present study, we use a NN with the objective of mapping the state vector of the thermoacoustic system to its time derivative; i.e., to obtain the differential equation that govern the thermoacoustic system. Such a NN which models the evolution of the dynamical system is termed as neural ODE by \citet{Chen_NODE_2018_7892}. A brief overview of NN and neural ODE along with training methodology is presented in the following subsection.  

\subsection{Mathematical description of neural network (NN)} 
Figure \ref{fig:nn_ode} shows the feed forward NN architecture used in this study. NN maps the input data to the output through series of mathematical manipulations. The layer of nodes that receives the input is the input layer and is denoted as Layer ($0$). There are $n^{(0)}$ inputs represented as vector ${Z}^{(0)} = [{z}^{(0)}_1,..,{z}^{(0)}_{n^{(0)}}]^T$. The layer of nodes that outputs the desired objective is Layer ($k$). There are $n^{(k)}$ outputs represented by the vector, ${Z}^{(k)} = [{z}^{(k)}_{1},..,{z}^{(k)}_{n^{(k)}}]^T$. The layers of nodes that lie between Layer ($0$) and Layer ($k$); i.e., layers $(1)$ to $(k-1)$, are called hidden layers. Biological neurons respond or generates action potential if the aggregate input stimuli is above a certain threshold. A similar behavior can be approximated in artificial NN at each node. A node maps the weighted sum of inputs to its output using a nonlinear mapping function called the activation function ($A$). There are different types of activation functions such as $sigmoid$, $tanh$, $ReLu$ \cite{Szandala2021}. Thus the output of the node $(j)$ in the Layer $(l)$; i.e., $z^{(l)}_j$ is computed as,
\begin{equation}
{z}^{(l)}_j=
A\bigg( \sum_{i=1}^{n^{(l-1)}} w^{(l)}_{ij} z^{(l-1)}_i\bigg).
\label{eq:node_output}
\end{equation} where, $w^{(l)}_{ij}$ is the weight assigned to the link that connects node $(i)$ in Layer $(l-1)$ to  node $(j)$ in layer $(l)$ and  $n^{(l-1)}$ is the number of nodes in Layer $(l-1)$. The vector of outputs of all nodes in layer $(l)$ is, $Z^{(l)} = [z^{(l)}_1,..,z^{(l)}_{n^{(l)}}]^T$. Thus, $Z^{(l)}$ can be represented as,
\begin{equation}
{Z}^{(l)}=
A\bigg( (W^{(l)})^{T} Z^{(l-1)}\bigg).
\label{eq:node_output_1}
\end{equation} where, $W^{(l)}=(w_{ij})\in \mathbb{R}^{n_{(l-1)}\times{n_{(l)}}}$, is the weight matrix. Defining a NN involves specifying the number of layers, nodes in each layer, activation function and weight matrices. This modular structure allows to modify the complexity of the function described by NN.

We denote the augmented matrix of all the weight matrices as, $\theta = [W^{(1)},..,W^{(k)}]$.
Thus, $\theta$ contains all the parameters of the NN. The NN can be expressed as a function ($f$) that maps the input vector $Z^{(0)}$ to the output vector $Z^{(k)}$,
\begin{equation}
    Z^{(k)} = f(Z^{(0)},\theta).
\end{equation}
For further discussion, we treat $X=Z^{(0)}$ as the input and $Y=Z^{(k)}$ as the output predicted by the NN. Thus, 
\begin{equation}
    Y = f(X,\theta).
\end{equation}
The data set obtained from experiment is, $\mathcal{D}=\{(X_i,Y_{tr,i})| i =(1,..,N)\}$, consisting of $N$ data samples, where $Y_{tr,i}$ is the true measured output from the experiment for the input $X_i$. Now, consider a NN which produces $Y_i$ as the output when the input is $X_i$. The penalty for failing to predict true $Y_{tr,i}$ and instead predicting $Y_i$ can be quantified using a loss function $L$, 
\begin{equation}
    L = L(Y_i,{Y_{tr,i}}) = L(f(X_i,\theta),Y_{tr,i}).
    \label{eq:loss}
\end{equation}
The average loss over the complete data set is,
\begin{equation}
    \mathcal{L} = \frac{1}{N} \sum_{i=1}^{i=N} L(f(X_i,\theta),Y_{tr,i}).
    \label{eq:loss_1}
\end{equation}
Training a NN involves optimizing the loss function $\mathcal{L}$ with respect to $\theta$. Based on the training objective, the appropriate loss function can be constructed \cite{janocha2017loss,bishop_2006}. The weight matrices corresponding to the different layers of the NN are randomly initialized before training. The process of optimization can be performed by gradient descend which involves marching the weights in the opposite direction of the gradient of the loss function with respect to $\theta$ \cite{Gradient_descent_Curry_1944}. The gradient is often computed using the back-propagation algorithm \cite{Rumelhart-1986}. Considering $\theta_n$ as the value of $\theta$ at $(n)^{th}$ iteration, during the $(n+1)^{th}$ iteration, the gradient decent algorithm updates this value of $\theta$ to $\theta_{n+1}$ as,
\begin{equation}
    \theta_{n+1} = \theta_{n} - \alpha \frac{\partial \mathcal{L}}{\partial \theta}\bigg|_{\theta_n} 
\end{equation} where, $\alpha$ is the learning rate. Training a NN involves minimizing the loss function with iterative application of this update rule. The discrepancy between the true (expected) output $Y_{tr}$ and the predicted output $Y_{pred}$ reduces upon repeated application of the update rule. Thus, trained or optimized NN can approximately predict the output.

\subsection{Mathematical description of neural ODE}
A dynamical system that is continuous in time can be represented in the form of a differential equation. NNs can be trained to approximate the differential equation of the dynamical system generating the data. This can be achieved by training NN to output the time derivative of the input state vector. As discussed before, the input vector as $X$ to the NN and the output vector as $Y$. This means that the output of NN, representing differential equation of the dynamical system, should output $Y$ which is approximately equal to time derivative of input state vector ${X}$; i.e., $Y\approx\dot{X}$. A NN that approximates the ordinary differential equation (ODE) of the dynamical system is termed as a neural ODE \cite{Chen_NODE_2018_7892}. Hence, the functional form of a neural ODE is,
\begin{equation}
    \dot{X} = f(X,\theta).
    \label{eq:NODE_diff_eq}
\end{equation}
The architecture of the neural ODE is the same as the previously discussed NN. Due to the constraint that the output is the time derivative of the input, the number of nodes in the output layer is the same as that of the input layer nodes. In Fig.\ref{fig:nn_ode}, variables in the braces of output layer forms the output vector of the neural ODE; i.e. $Y=[\dot{z}_1^{(0)}, \dot{z}_2^{(0)}, ..., \dot{z}_{n^{(k)}}^{(0)}]= \dot{X}$. The evolution of the dynamical system with time can be obtained by integrating the neural ODE numerically. When $X_{tr}(t_0)$ is the true initial state of the system at time $t_0$ (known from the data), the neural ODE predicts  the the state of the system as $X(t_i)$ at time $t_i$ as, 
\begin{equation}
     {{X}} (t_i) = {X}_{tr}(t_0)+ \int_{t_0}^{t_i}f({X}(t),{\theta}) dt
\end{equation}
If, $X_{tr}(t_i)$ is the true state of the system at time $t_i$, then the loss incurred due to prediction $X(t_i)$ is, 
\begin{equation}
    L = L(X(t_i),X{tr}(t_i)).
    \label{eq:loss_2}
\end{equation}
Optimization of the loss function with respect to $\theta$ can be performed using the adjoint optimization method, details of which can be found in \cite{Chen_NODE_2018_7892}. The parameter vector $\theta$ obtained after training is characteristic to the dynamical state of the system.

\subsection{Model description}{\label{subsec:model_description}}
A set of variables $(x_1, x_2,.., x_d)$ that gives the complete description of the state of system are known as state variables. Here, $x_i$ is a state variable and there are total $d$ state variables. The $d$ dimensional space with state variables as coordinates is referred to as the phase space. Thus, the phase space contains all possible states that the system can assume. In practical systems only a finite number of state variables can be measured. Under this condition, in order to visualize and analyze the asymptotic dynamics of an experimental system, we can reconstruct a phase space according to delay embedding theorem proposed by \citet{Takens1981}. For thermoacoustic systems the unsteady acoustic pressure, $p'(t)$ can easily be measured. The evolution of the thermoacoustic system can be visualized by reconstructing the phase space using delay-coordinates. The delay coordinate can be expressed as, $[p'(t), p'(t+\tau), .., p'(t+(d-1)\tau)]$. Here, $\tau$ is the delay and $d$ is the dimension of the reconstructed phase space known as the embedding dimension. The dimension of the reconstructed phase space should be sufficiently high to avoid overlap of the trajectories and false neighbors. Reconstruction of phase space involves two steps 1) finding the optimal time delay $\tau$ and 2) finding the dimension of the phase space \cite{abarbanel_1993, kantz_schreiber_2003}. For the present analysis we used the method of average mutual information (AMI) to obtain $\tau$. The method has its roots in the information theory. The optimal delay is chosen such that the original time series $[p'(t_1),p'(t_2),..]$ conveys the least information about the delay time series $[p'(t_1+\tau),p'(t_2+\tau),..]$. 

We used the method of false nearest neighbors (FNN) to obtain the embedding dimension. The co-ordinates of the dynamical system at a time instance $t_i$ with embedding dimension $d$ is $y_d(t_i)=[p'(t_i), p'(t_i+\tau),.., p'(t_i + (d-1)\tau)]$. The $d$ for which the total number of FNN drops to zero is chosen as the embedding dimension. A brief description of AMI and FNN is given in Appendix \ref{app:AMI_FNN}. 
The state vector for the reconstructed phase space using the time series of pressure is, $X(t) = [p'(t), p'(t+\tau),..,p'(t+(d-1)\tau)]$. The dynamical system is then represented as:
\begin{equation}
    \frac{d{X}(t)}{dt} = {f}({X}(t),t,{\theta})
\end{equation} 
This equation is the same as equation (\ref{eq:NODE_diff_eq}), where ${f}$ is the function representing a NN. We construct NN with $tanh$ activation function and one hidden layer. According to universal approximation theorem, a NN even with a single hidden layer  with a smooth activation function can approximate any function if sufficiently hidden units or nodes are considered \cite{Kur_1989_Universal_Approx}. The $tanh$ activation function is used as it is smooth, changes sign across the zero and could approximate smooth functions better compared to piecewise linear functions \cite{Szandala2021}. The absolute value of the error is used to construct the loss function ($L$) as:
\begin{equation}
    L({X}(t_i))=|{X}(t_i) -{X}_{tr}(t_i)|.
    \label{eq:loss_NODE_exp_1}
\end{equation}
During training, the absolute value of the average loss function is optimized with respect to the neural ODE parameter vector $\theta$ at a fixed Reynolds number ($Re$). Thus the parameters of the neural ODE depends on the Reynolds number; i.e., $\theta = \theta (Re)$. The neural ODE trained on the experimental data gives an approximate dynamical system representation for the experimental system.

\subsection{Precursor measure}{\label{subsec:model_pre}}

Thermoacoustic systems are characterized by the mutual interaction of the heat source and the acoustic field in the confinement. Pressure fluctuations inside the combustor can easily be monitored using a microphone. Measurements of heat release rate fluctuations are often unavailable due to inaccessibility of the combustor to photo-multiplier tube. Thus, one should be able to predict the onset of periodic oscillations using the time series of pressure fluctuations alone. We reconstruct the phase space using the time series of pressure as, $X_p(t) = [p'(t),p'(t+\tau_p),..,p'(t+(d_p-1)\tau_p)]$. Here, $\tau_p$ and $d_p$ are the time delay and embedding dimension respectively, for the time series of pressure fluctuations. The dynamical system is then represented as:
\begin{equation}
    \frac{d{X_p}(t)}{dt} = {f}({X_p}(t),t,{\theta}).
\end{equation} 
The construction of loss function and its optimization are already discussed in \ref{subsec:model_description}. The present model constructed based on the time series of pressure fluctuations has parameter vector $\theta$ that governs the behavior of the model. A smooth variation of these parameters can lead to a qualitative change in the dynamical state of the system known as bifurcation.

The time series of $p'$ with $N$ data points is measured for each operating state defined by the $Re$ of the flow. We train our NN on the first $n_w$ continuous data samples. After training with this window of $n_w$ data points, the window is shifted by $n_s$ data points and the neural ODE is trained again with the new $n_w$ data points. Therefore, for each operating state, $[N/n_s]$ different neural ODEs and their corresponding parameter vectors $\theta$ are obtained ([.] is the greatest integer function). To get single representative $\theta$ for a given $Re$, we average the parameter vector to get $\bar{\theta}(Re)=[\overline{{W}}^{(1)},..,\overline{{W}}^{(k)}]$. 

A sine wave with a frequency ($f$ Hz) equal to the limit cycle oscillations ($f_{LC}$ Hz) is constructed (we observed that for 5\% deviation in $f$ from $f_{LC}$, the proposed precursor measures shows the expected behavior). The neural ODE is trained on this sine wave and the obtained parameter vector, $\theta_{ref}=[{W}^{(1)}_{ref},..,{W}^{(k)}_{ref}]$ is used as a reference. The cosine of  the angle between the ${{W}}^{(l)}(Re)$ and ${{W}}^{(l)}_{ref}$ quantifies the alignment of the weight matrix of the current operating state at specified $Re$ with the reference weight matrix obtained from sine wave. This precursor measure $m(Re)$ is obtained as,
\begin{equation}
    m^{(l)}(Re)= \frac{<{{W}}^{(l)}(Re),{{W}}^{(l)}_{ref}>} {|\overline{{W}}^{(l)}(Re_0)|_2|{{W}}^{(l)}_{ref}|_2}.
    \label{eq:precursor}
\end{equation} where, $<,>$ denotes the inner product and $|.|_2$ denotes the $L_2$ norm of the matrix.

\subsection{Modelling the thermoacoustic system}{\label{subsec:model_HR}}

 For turbulent combustors the reacting flow field is the source of heat. Modeling a thermoacoustic system using linearized wave equation needs a model for the heat source. The response of the heat release rate to velocity fluctuations is often modeled using a flame transfer function or a flame describing function. The flame transfer function models the heat release rate as a linear response to the velocity fluctuations and hence the gain and phase of the transfer function are independent of the amplitude of the acoustic velocity fluctuations. However, the flame response is a nonlinear function of the amplitude of the acoustic velocity fluctuations. To capture this nonlinear dependence, researchers use a flame transfer function with dependence on acoustic fluctuation amplitude known as the flame describing function. In general, the heat release rate is modulated by the oscillations in the flame surface, equivalence ratio, acoustic velocity and pressure fluctuations. The net heat release rate can be obtained by integrating the heat release rate over the flame surface.
 The heat release rate can, in general, expressed as,
\begin{equation}
    \dot{q}=\dot{q}(u',p')
\end{equation}where, ${u}'$ and ${p}'$ are the velocity and pressure fluctuations. In present study, we are using the net heat release rate ($\dot{q}$) and not the mean subtracted ($\dot{q}'$).

In present study, the time series of pressure ($p'$) and heat release rate fluctuations ($\dot{q}$) are obtained from experiments conducted on a turbulent combustor. Measurements of the velocity fluctuations $u'$ are absent. As discussed before, the reconstruction of phase space can be used to create a high dimensional phase space to describe the dynamics. Thus, dynamical system representation can be obtained for the thermoacoustic system, even in the absence of measurement of $u'$. To reconstruct the phase space, the delay and the embedding dimension for the $p'$ and $\dot{q}$ are individually obtained as ($\tau_p$, $d_p$) and ($\tau_q$, $d_q$). The state vector for the reconstruction of phase is constructed as, $X_{pq}$ = $[p'(t),p'(t+\tau_p),..,p'(t+(d_p-1)\tau_p), \dot{q}(t),\dot{q}(t+\tau_q),..,\dot{q}(t+(d_q-1)\tau_q)]$. Thus, thermoacoustic system can be represented as,
\begin{equation}
    \frac{d{X_{pq}}(t)}{dt} = {f}({X_{pq}}(t),t,{\theta})
\end{equation} 
We use neural ODE to approximate this mapping of state vector to its time derivative. On training the neural ODE on time series of $p'$ and $\dot{q}$, we get a model for the thermoacoustic system. 
The absolute value of the difference between the predicted and the true time series is used as loss function. Optimization of the average loss function with $\theta$ gives the parameter vector for the neural ODE. Since the neural ODE is trained on time series representing the simultaneous variation of $p'$ and $\dot{q}$, a NN with the optimum $\theta$ can represent the coupled behavior of the unsteady heat release rate and acoustic perturbations in a thermoacoustic system.
\section{Experiments}\label{sec:experiments}
\begin{figure}[ht!]
    \centering
         \includegraphics[width=0.9\textwidth]{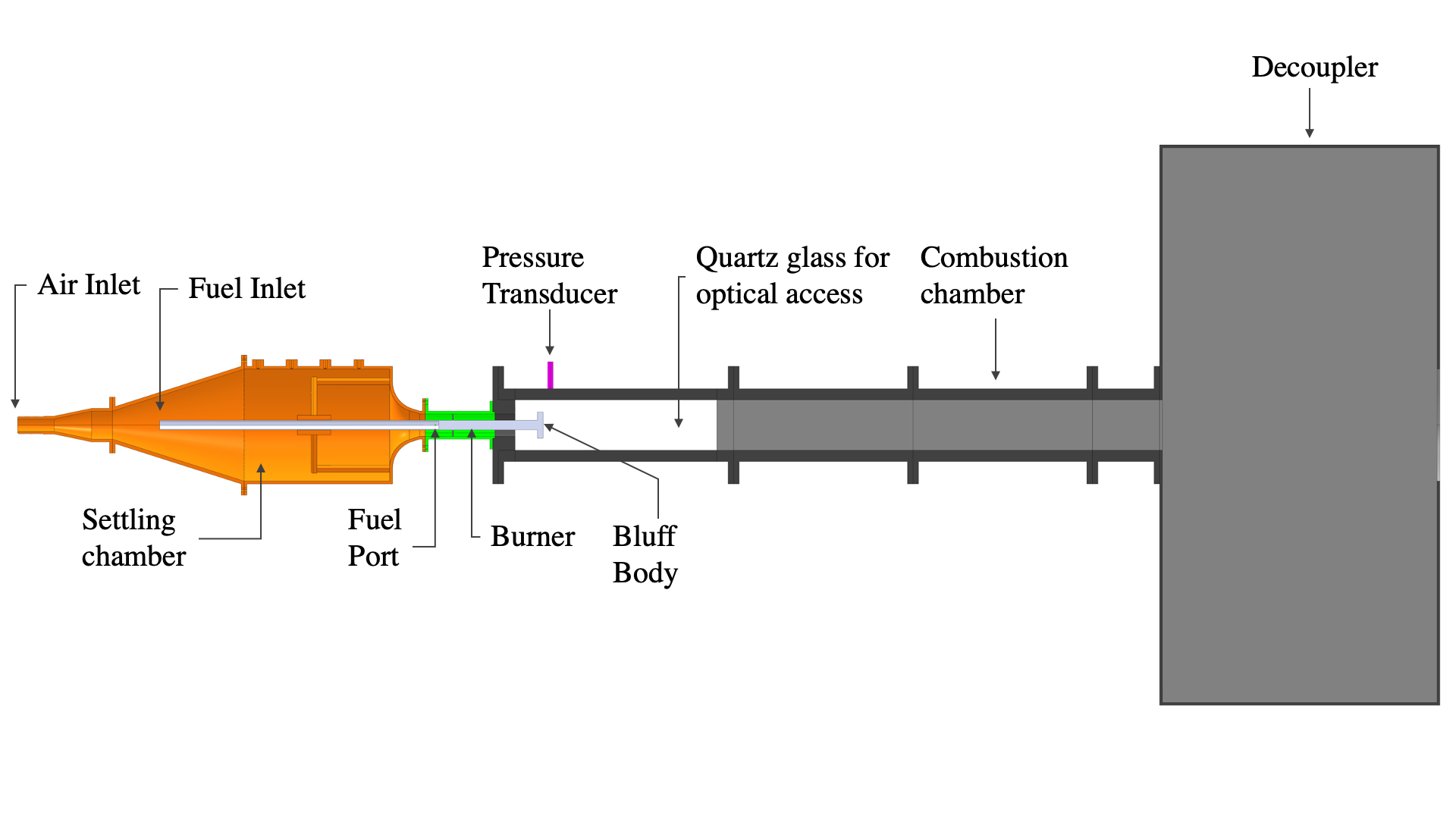}
     \caption{The sectional view of an experimental setup of the turbulent combustor with a bluff body used as flame holder. The combustor consists of four main units 1) a settling chamber, 2) a burner, 3) a combustion chamber, and 4) a decoupler. Air flow coming from the air inlet passes through the setting chamber before entering into the combustion chamber. The cylindrical bluff body protrudes into the combustion chamber. The pressure transducer is mounted on the side wall of the combustor, next to the bluff body. The two opposite walls of the combustion chamber are replaced with quartz glass to enable optical access. The photo-multiplier tube (PMT) (not shown in the figure) is mounted with its axis normal to the quartz glass. The exit of the combustion chamber is connected to the decoupler.}
     \label{fig:exp_setup_TA}
\end{figure}  

A schematic of the turbulent combustor is shown in the  Fig. \ref{fig:exp_setup_TA}. The details of the experiments are reported by \citet{unni_sujith_2015}. 
The experiments are preformed with a bluff body as flame holding mechanism. The combustor length is 1400 mm. Liquefied petroleum gas (LPG: butane $40 \%$ and propane $60\%$) is used as fuel. The air is partially premixed with the fuel before the mixture enters the combustion chamber. Mass flow controllers (Alicat MCR series) are used to control air and fuel flow with an uncertainty of $\pm$ ($0.8\%$ of reading + $0.2\%$ of full scale). The different mass flow rates of reactants through the combustor are obtained by changing the air flow rate in steps from 448 slpm to 878 slpm in quasi-steady manner while maintaining a constant fuel flow rate of 28 slpm.  The corresponding Reynolds number of the flow varies from $Re = 1.03\times 10^4$ to $Re = 2.03 \times 10^4$, with an uncertainty of 2.7$\%$, during the experiment. The reader may refer to \citet{unni_sujith_2015} for the details of the measurements.

A piezoelectric transducer (PCB103B02 transducer sensitivity $217.5$ mV/kPa with uncertainty $\pm0.15$ Pa) is used to measure the time series of pressure fluctuations ($p'$). The pressure transducer is mounted at the pressure antinode near the backward facing step. The unsteady heat release rate ($\dot{q}$) is measured using a photo multiplier tube (PMT; Hamamatsu H10722-01). The PMT is mounted near the bluff body with axis normal to the quartz glass. For each $Re$, the time series of pressure $(p')$ and heat release rate ($\dot{q}$) fluctuations are recorded for $3$ s at a sampling rate of 10 kHz using an A-D card (NI-6143, 16 bit). 


\section{Results and discussion}\label{sec:results_discussion}

The different dynamical states  of the turbulent combustor are achived by varying the air flow rate in steps and maintaining the fuel flow rate constant. The time series of pressure fluctuations for incremental $Re$ are stacked and an augmented time series is shown in the Fig. \ref{fig:res_TA} (a). We can observe that the time series of pressure fluctuations is chaotic having a small amplitude during the state shown in Fig. \ref{fig:res_TA} (b) for which $Re=1.07\times 10^4$. On increasing $Re$, the time series shows bursts of periodic oscillations amidst epochs of aperiodicity as shown in Fig. \ref{fig:res_TA} (c) which corresponds to $Re = 1.32\times 10^{4}$. Further increase in $Re$ leads to thermoacousic instability with ordered periodic high amplitude pressure oscillations. Figure \ref{fig:res_TA} (d) shows the state of thermoacoustic instability at $Re = 1.92 \times 10^4$. Thus, the time series data obtained for thermoacoustic system shows a transition from chaos to periodic oscillations through the state of intermittency.  

\subsection{Thermoacoustic system modeling}
{\label{subsec:TA_Model}}

A thermoacoustic system involves the interaction between the acoustic field and the heat the release rate. Thus, we use $X_{pq}$ as the state vector as discussed in Sec. \ref{subsec:model_HR}.

The NN considered for the present analysis, contains only a single hidden layer with $tanh$ activation function. For the present analysis, the hidden layer always contains 100 nodes (the convergence of the predicted time series to the experimental data is given in Appendix \ref{subsec:conv_neural_ODE_hid}). The input and output layer contains nodes equal to the dimension of the state vector.

\begin{figure}[ht!]
     \centering
         \centering
         \includegraphics[width=0.9\textwidth]{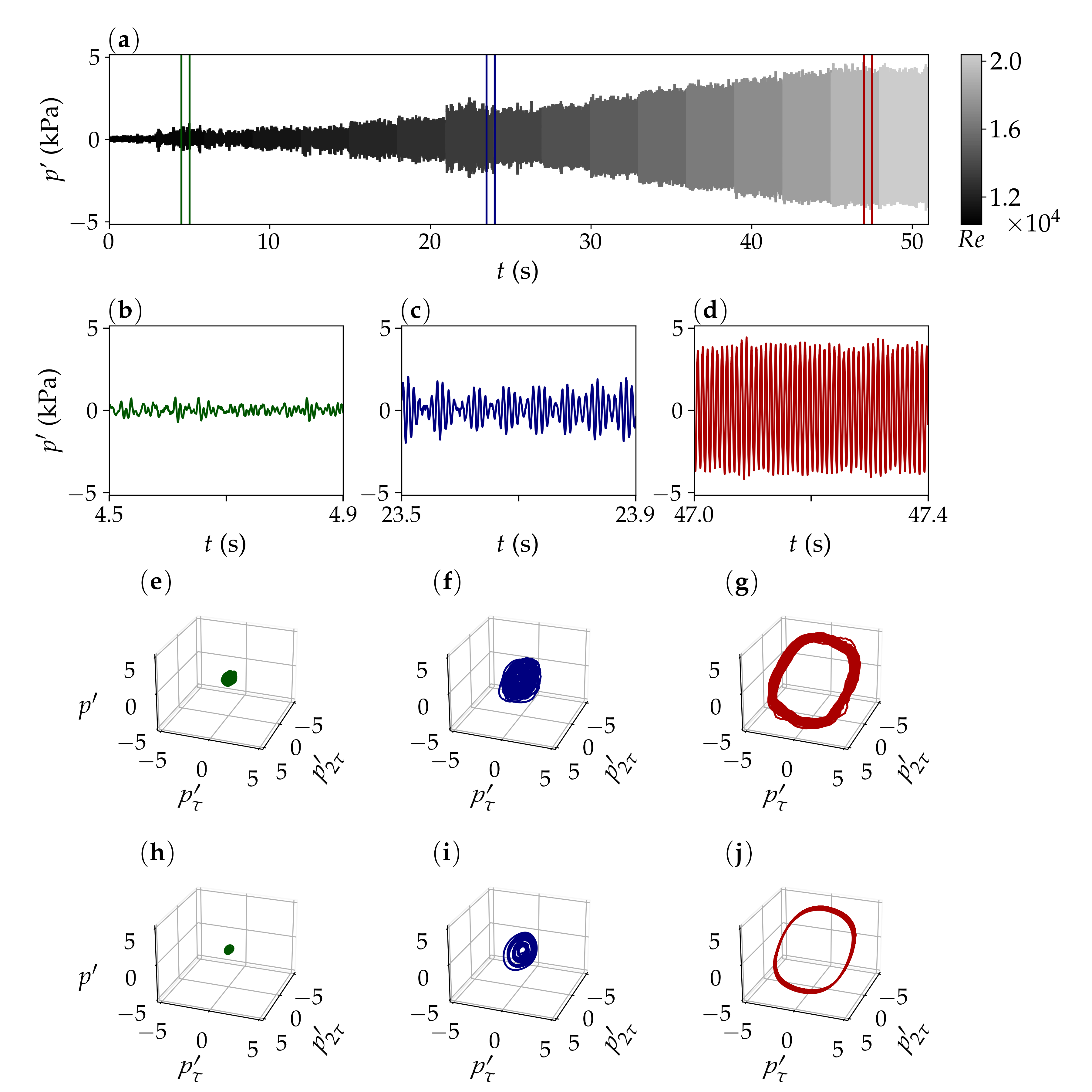}
     \caption{Variation in time series of pressure fluctuations as the thermoacoustic system in the turbulent combustor transitions from the state of combustion noise (chaotic) to thermoacoustic instability (limit cycle oscillations). a) For illustration purpose, the time series of pressure fluctuations (each $3~s$ long) obtained for a sequence of $Re$ values, varied in a quasi steady manner, are stacked together to make a continuous time series representing the transition. (b), (c) and (d) shows zoomed in view of $p'$ at $Re$ = $1.07\times10^4$,  $1.32\times 10^4$, $1.92\times10^4$ respectively. (e), (f) and (g) shows the reconstructed phase portrait of the experimental time series shown in (b), (c) and (d) respectively in ($p'(t)$, $p'(t+\tau)$, $p'(t+2\tau)$) coordinates. (h), (i) and (j) shows the trajectory in phase space obtained using a neural ODE with the the same initial condition as that of phase trajectory in (e), (f) and (g) respectively.}
     \label{fig:res_TA}
\end{figure}  
 To visualize the change in the dynamical state of the system, all these time series, obtained for a sequence of $Re$, are stacked and are shown in Fig. \ref{fig:res_TA}(a). We train the neural ODE on the first $n_w$ data points. To compare this length of data with the number of data points $n_c$ in one period of limit cycle oscillation, we use $N_c = n_w/n_c$. We used $N_c\approx48$ to train the network. We shift the training window by $n_s$ data points to train another NN. Thus, for each $Re$ we have $[N/n_s]$ parameter vectors. As discussed, to get the single representative neural ODE parameter vector, the average parameter vector $\bar{\theta}(Re)$ is calculated. The training is performed with a fixed number of $2000$ iterations and a learning rate of $\alpha=10^{-2}$.  

\begin{figure}[ht!]
	\centering
    \includegraphics[width=0.85\textwidth]{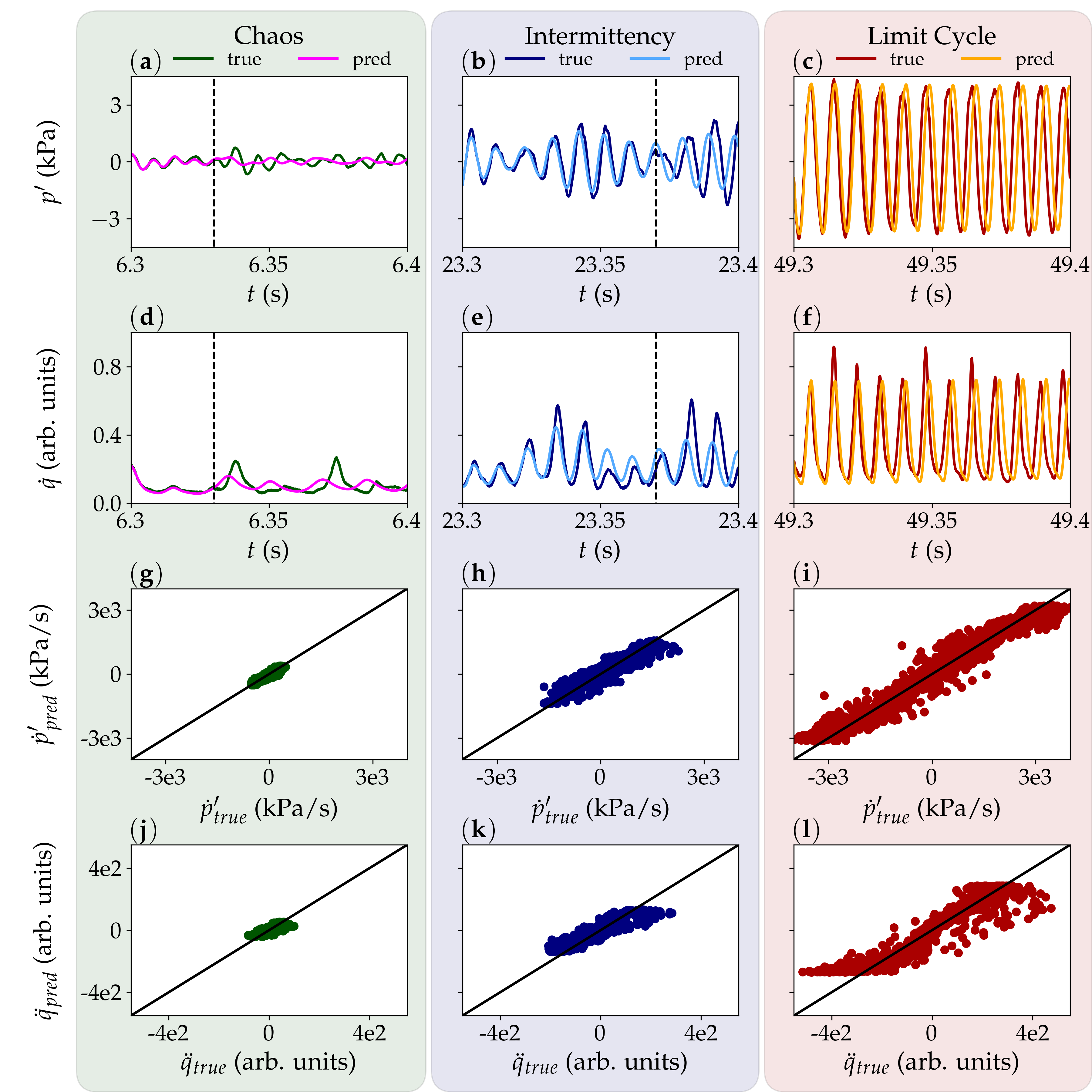} 
    \caption{Comparison of true time series of measured state variables ($p'$ and $\dot{q}$) with the corresponding time series predicted using neural ODE. Different dynamical states are shown in different columns. The left column shows the state of chaos, the middle column shows the state of  intermittency and  the right column shows the state of limit cycle oscillations. Figures (a), (b) and (c) show the time series of $p'_{true}$ measured from the experiment and $p'_{pred}$ predicted using neural ODE. Figures (d), (e) and (f) compare the time series of $\dot{q}_{true}$ measured from the experiment and $\dot{q}_{pred}$ predicted using neural ODE. The dashed line in (a), (b), (d), (e) indicates the time instance after which deviation grows. Figures (g), (h) and (i) show comparison of $\dot{p}'_{true}$ $=$ $(d{p}'/dt)_{true}$ and $\dot{p}'_{pred}$ $=$ $(d{\dot{p}}'/dt)_{pred}$ predicted using neural ODE. Figures (j), (k) and (l) compare $\ddot{q}_{true}$ $=$ $(d\dot{q}/dt)_{true}$  and $\ddot{q}_{pred}$ $=$ $(d\dot{q}/dt)_{pred}$, measured from the experiment and predicted using the neural ODE.}
    \label{fig:test_comp}
\end{figure} 

The phase space reconstruction is performed using $AMI$ and $FNN$ methods as discussed in Sec. \ref{subsec:model_HR}. The time series data of periodic oscillations is used to obtain $AMI$ and $FNN$. The variation of the $AMI$ with delay ($\tau$) and $FNN$ with embedding dimension $d$ are shown in Fig. \ref{fig:delay_emb_TA} in Appendix \ref{app:AMI_FNN} . For the time series of $\dot{q}$ and $p'$, the $AMI$ attains a first minima at $\tau_q$ = 22 and $\tau_p$ = 21 respectively; these values ($\tau_q$ and $\tau_p$) are chosen as optimal delay. For time series of $\dot{q}$ and $p'$ the number of $FNN$ become zero at dimension $d_q=10$ and $d_p=7$ respectively. The combined dimension $d= d_q+d_p$ is chosen as the embedding dimension to reconstruct the phase space.

The phase space trajectory of thermoacoustic system is reconstructed in ($p'(t)$, ${p}'(t+\tau)$, ${p}'(t+2\tau)$) coordinates for visualization. The phase space trajectory constructed from experimental data for the state of chaotic, intermittent and limit cycle oscillations are shown in Fig. \ref{fig:res_TA} (e), (f) and (g) respectively. These time series are reconstructed with the neural ODE and are shown in Fig. \ref{fig:res_TA} (h), (i) and (j) respectively. This reconstruction is performed by using the same parameter ${\theta}(Re)$ and the initial condition as the corresponding time series obtained from experiments shown in Fig. \ref{fig:res_TA} (e), (f) and (g) respectively. Integration is performed with an explicit $4^{th}$ order Runge-Kutta integrator \cite{dopri_5} 
for the duration of 0.4 s.  Considering time period of limit cycle oscillations as $T_{LC}$, we integrate the neural ODE for close to $48T_{LC}$ duration. 

For the experimental data, the phase space trajectory for the chaotic state shows that there is a cluttered trajectory, occupying only a small region in the phase space as shown in Fig. \ref{fig:res_TA} (e). This figure also shows that the trajectory does not settle to a fixed point. The same behavior can be observed for the reconstructed phase trajectory using the neural ODE shown in Fig. \ref{fig:res_TA} (h). For the state of intermittency, Fig. \ref{fig:res_TA} (i),the reconstructed phase space trajectories using neural ODE shows different orbits indicating switching of the system dynamics between chaos and periodic oscillations which is not clearly visible in Fig. \ref{fig:res_TA} (f) due to overlapping of multiple bursts of periodic oscillations with slightly different amplitude. Fig. \ref{fig:res_TA} (g) shows that for the state of limit cycle oscillations the trajectory forms a closed loop for experimental data and matches well with the reconstructed phase space trajectory obtained using neural ODE which is shown in Fig. \ref{fig:res_TA} (j). Considering that the integration is performed for $48T_{LC}$ the phase space trajectories show that the neural ODE could predict the asymptotic behavior for the states of chaos, intermittency and limit cycle oscillations. Thus, Fig. \ref{fig:res_TA} clearly shows that the neural ODE performs well on the training data. 

The neural ODE is tested with initial condition outside the training data, and the comparison of time series is shown in Fig. \ref{fig:test_comp}. The neural ODE is integrated for a duration $\approx12T_{LC}$. Figure \ref{fig:test_comp} (a-c)  and \ref{fig:test_comp} (d-f) respectively compares the true and predicted time series of pressure $p'$ and heat release rate $\dot{q}$ fluctuations for the state of chaos, intermittency and limit cycle oscillations. These dynamical states corresponds to the $Re$ = $1.07\times10^4$,  $1.32\times 10^4$, $1.92\times10^4$ respectively. For chaos, the predicted trajectories for $p'$ and $\dot{q}$ follows the true trajectory for $\approx 3T_{LC}\approx 3.3 T_{\lambda}$ (here $T_{\lambda}$ is the Lyapunov time scale) which is marked by dashed vertical line in the figure and after that deviation grows. For the state of intermittency, the predicted and true trajectories for $p'$ matches well for $\approx8T_{LC}$ and indicated by dashed vertical line in the figure. After that the neural ODE qualitatively captures the true time series. For the state of limit cycle oscillations, comparison of predicted and true time series of  $p'$ and $\dot{q}$ shows that predictions agree well with true data. As the neural ODE predicts the time derivative of the state variables, Fig. \ref{fig:test_comp} (g-i) and (j-l) shows the comparison of true and predicted time derivative of $p'$ and $\dot{q}$ (i.e., $\dot{p}'$ and  $\ddot{q}$) for the state of chaos, intermittency and limit cycle oscillations, respectively. The time derivative of the experimental data is obtained using a $2^{nd}$ order central difference scheme. Ideally, if the true and predicted values match; i.e., $((.)_{true}=(.)_{pred})$ then the point should fall on the reference diagonal line as shown in Fig. \ref{fig:test_comp} (g-l). For all the three states considered here, all the points fall in the close vicinity of the diagonal line.
Thus, neural ODE captures the dynamics of the thermoacoustic system for all the three states and performs well over testing data.

\subsection{Precursor to thermoacoustic instability} {\label{subsec:Precursor_TAI}}
\begin{figure}[h]
     \centering
        \includegraphics[width=\textwidth]{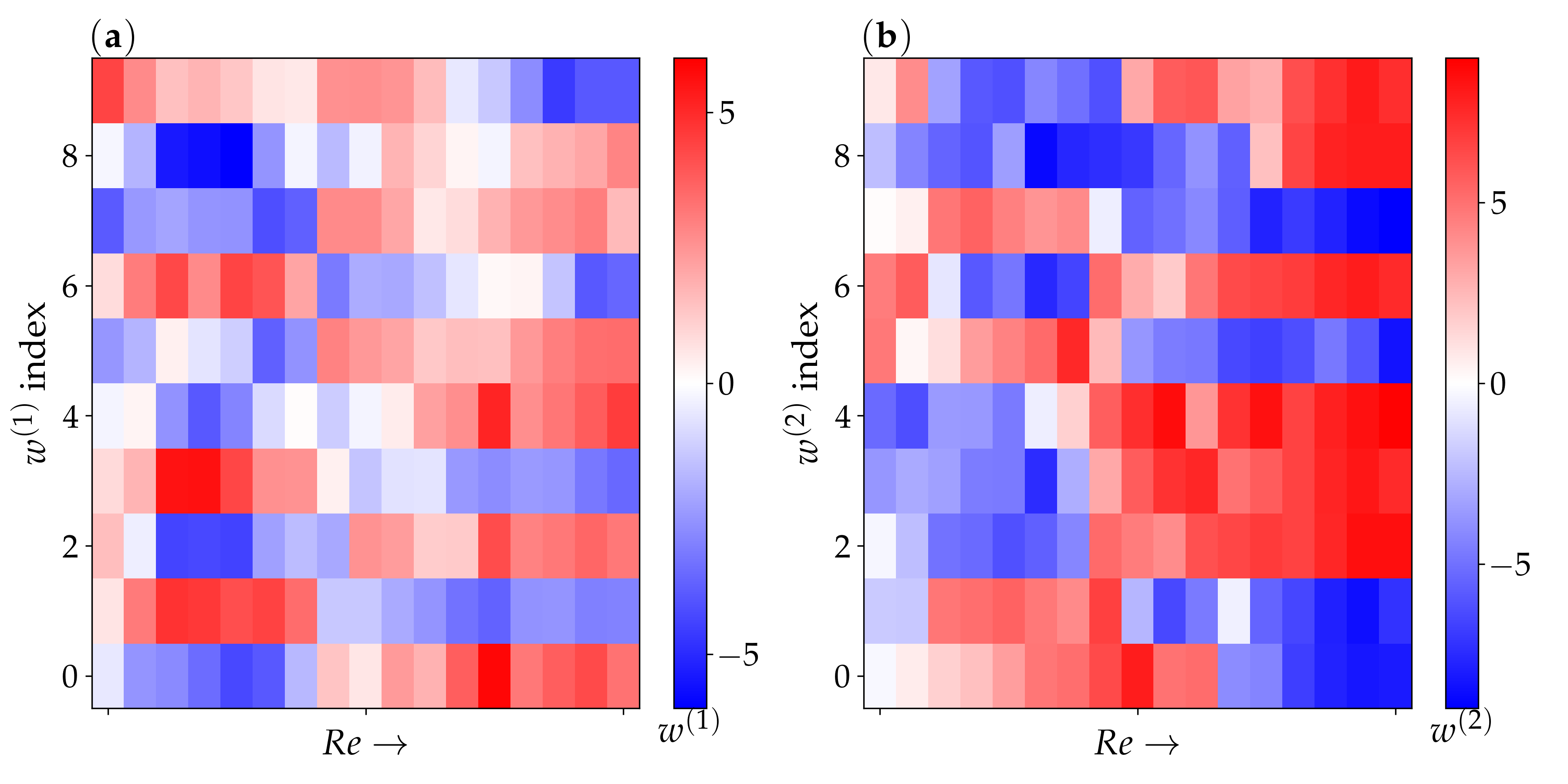}
     \caption{Evolution of the ten elements of weight matrix with Reynolds number for thermoacoustic system. These ten elements are chosen from each weight matrix with highest variance. (a) Evolution of ten elements from $W^{(1)}$ matrix,(b) Evolution of ten elements of $W^{(2)}$, with highest variance. Color bar indicates the value of weight matrix elements.}
     \label{fig:res_ta_color_map}
\end{figure}  

\begin{figure}[h]
     \centering
        \includegraphics[width=\textwidth]{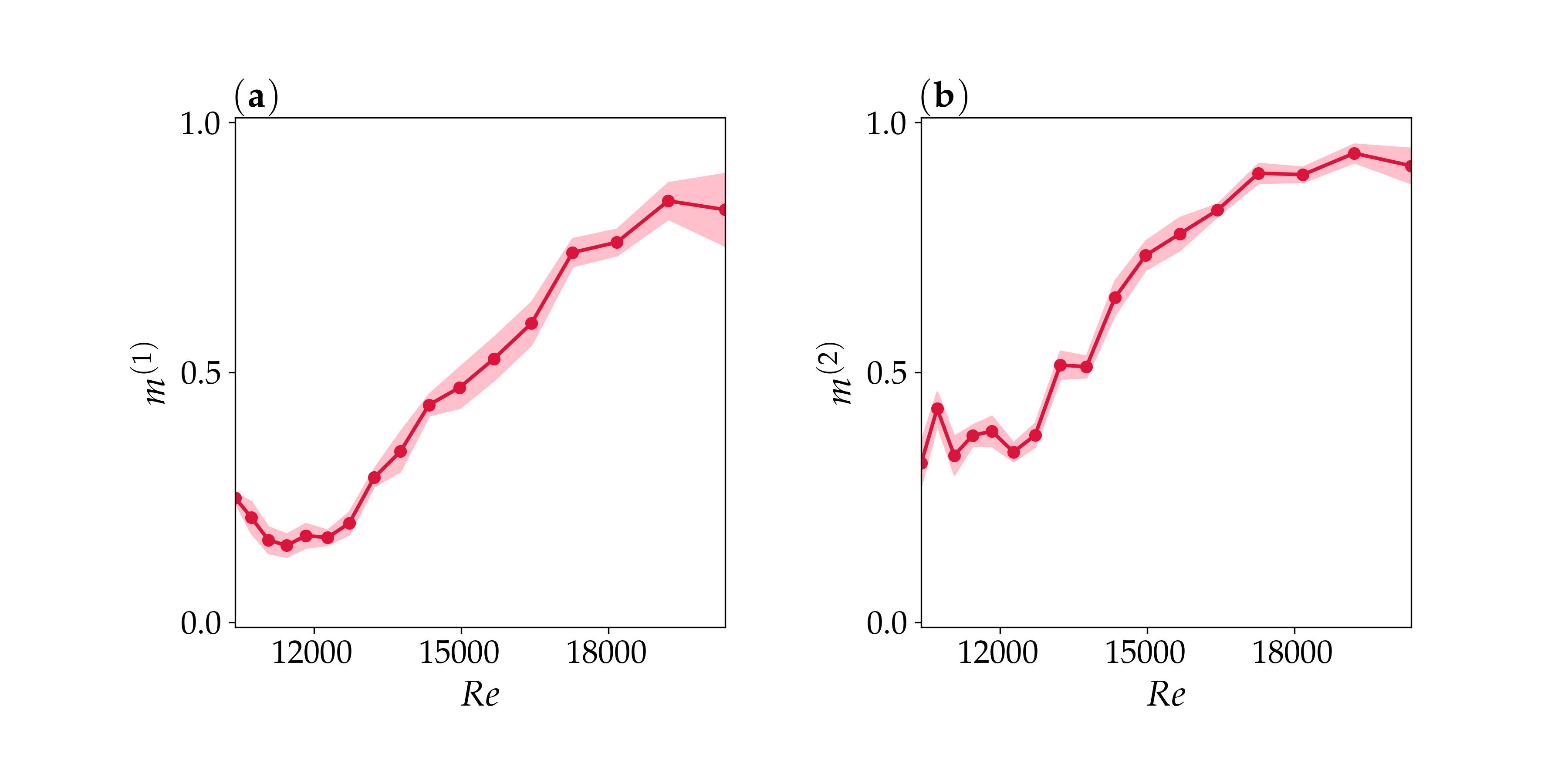}
     \caption{Variation of the normalized distance $m^{(l)}$ with Reynolds number for thermoacoustic system. (a) Variation of $m$ for the first weight matrix ($W^{(1)}$), (b) Variation of $d$ for  second weight matrix ($W^{(2)}$). The dots denotes the normalized distance of mean weight matrix ($\bar{m}^{(l)}$) from the reference state. The color band shows the region of $\pm$ (standered deviation)}
     \label{fig:res_precursor}
\end{figure} 

For turbulent combustor, thermoacoustic instability is presaged by the formation of coherent structures, intermittency and a reduction in the dimensionality as discussed in Sec. \ref{sec:introduction}. For dynamical systems, smooth variation of the parameters of the differential equation can alter the dynamical state of the system. Such a bifurcation could also occur in the neural ODE on changing the parameter vector $\theta$. The dynamics of thermoacoustic system changes with $Re$; hence $\theta$ should be a function of $Re$ ; i.e., $\theta=\theta(Re)$. Now, we create a parametric space where the parameter vector $\theta$ is the position vector for the state of the thermoacoustic system. The position of the thermoacoustic system in parametric space changes on changing the $Re$. We choose ten elements from each weight matrix of the NN to visualize the evolution of elements of weight matrix with $Re$. The variation of these elements with $Re$ for $W^{(1)}$ is shown in the Fig. \ref{fig:res_ta_color_map} (a) and for $W^{(2)}$ is shown in Fig. \ref{fig:res_ta_color_map} (b). For the state of chaos with small $Re$, some elements of weight matrix $W^{(1)}$ shows high magnitude. As the thermoacoustic system approaches state of limit cycle oscillations by increasing $Re$, the magnitude of the weights approaches to zero. On the other hand weights from $W^{(2)}$ are close to zero for the state of chaos and the magnitude of the weights increases as the thermoacoustic system approaches limit cycle oscillations. This observed pattern shows that weight matrices hold the information about the proximity of the thermoacoustic system to the state of limit cycle oscillations.

We analyze the evolution of the thermoacoustic system in the parametric space with the elements of the parameter vector $\theta$ as coordinate. The point that describes the position of the thermoacoustic system in parametric space moves closer to the point that describes the reference state, obtained using  sinusoidal signal, on increasing the $Re$. As the point describing the dynamic state of the system in the parametric space approaches the reference state, the parameter vector of the system becomes aligned to the reference vector. To quantify the proximity of the thermoacoustic system to the onset of limit cycle oscillations we propose a measure $m^{(l)}$ as discussed in Sec. \ref{subsec:model_pre}. As discussed before, we have $[N/n_s]$ parameter vectors for each operating state. On evaluating $m^{(l)}$ for all $[N/n_s]$ parameter vectors we compute the average measure $\bar{m}^{(l)}(Re)$ and the standard deviation $\sigma(Re)$. Thus, for the sequence of Reynolds numbers ($Re$) we have  $\bar{m}^{(l)}(Re)$ and $\sigma(Re)$. The variation of $m^{(l)}$ with $Re$ is shown in Fig. \ref{fig:res_precursor}. The marker shows the mean value of the measure $\bar{m}^{(l)}(Re)$ and $\pm \sigma(Re)$ bound is shown using the color band. The figure includes plots for variation of $m^{(1)}$ and $m^{(2)}$ that  corresponds to the weight matrix $W^{(1)}$, $W^{(2)}$ respectively. Both the figures show that $m$ approaches a value of one as the system approaches the state of limit cycle oscillations.  Thus, we propose that $m$ could be used as a measure to quantify the proximity of the dynamical state of the thermoacoustic system to the onset of thermoacoustic instability. To assess the generality of the proposed framework to detect the proximity of the system dynamics to oscillatory instability, we apply this methodology to an aeroacoustic system with turbulent flow as described in Appendix \ref{res_generlizability}. 

\section{Conclusion}\label{sec:conclusion}


Using the neural ODE to represent the phase space dynamics, we obtained a data-driven model for turbulent thermoacoustic systems which could qualitatively capture the states of chaotic, and intermittent oscillations and quantitatively captures the state of limit cycle oscillations. Reconstructing the phase space using delay embedding we demonstrated how the dynamical system can be identified using neural ODE when data is available for only a few state variables such as $p'$ and $\dot{q}$. Using neural ODE to model thermoacoustic system provides an alternative data-driven methodology for modeling the thermoacoustic system, instead of having to explicitly model the functional relationship between the state variables. The model also provides flexibility to incorporate additional state variables if their time series are available.

We also proposed a precursor measure $m^{(l)}$, defined using the parameters of neural ODE, which approaches a value of one during the onset of thermoacoustic system. We also showed the the generality of the proposed measure to detect the onset of periodic oscillations by detecting the onset of aeroacoustic instabilities.

\begin{acknowledgments}
Jayesh Dhadphale is indebted to Ministry of Human Resource Development for providing the fellowship under The Prime Minister's Research Fellows (PMRF) scheme. R. I. Sujith acknowleges the funding from the Science and Engineering Research Board (SERB) of the Department of Science and Technology (DST) through a J. C. Bose Fellowship (No. JCB/2018/000034/SSC) and from the IOE initiative (SB/2021/0845/AE/MHRD/002696).
\end{acknowledgments}

\appendix

\section{Appendixes}
\subsection{Phase space reconstruction}\label{app:AMI_FNN}

\begin{figure}[t!]
     \centering
         \includegraphics[width=\textwidth]{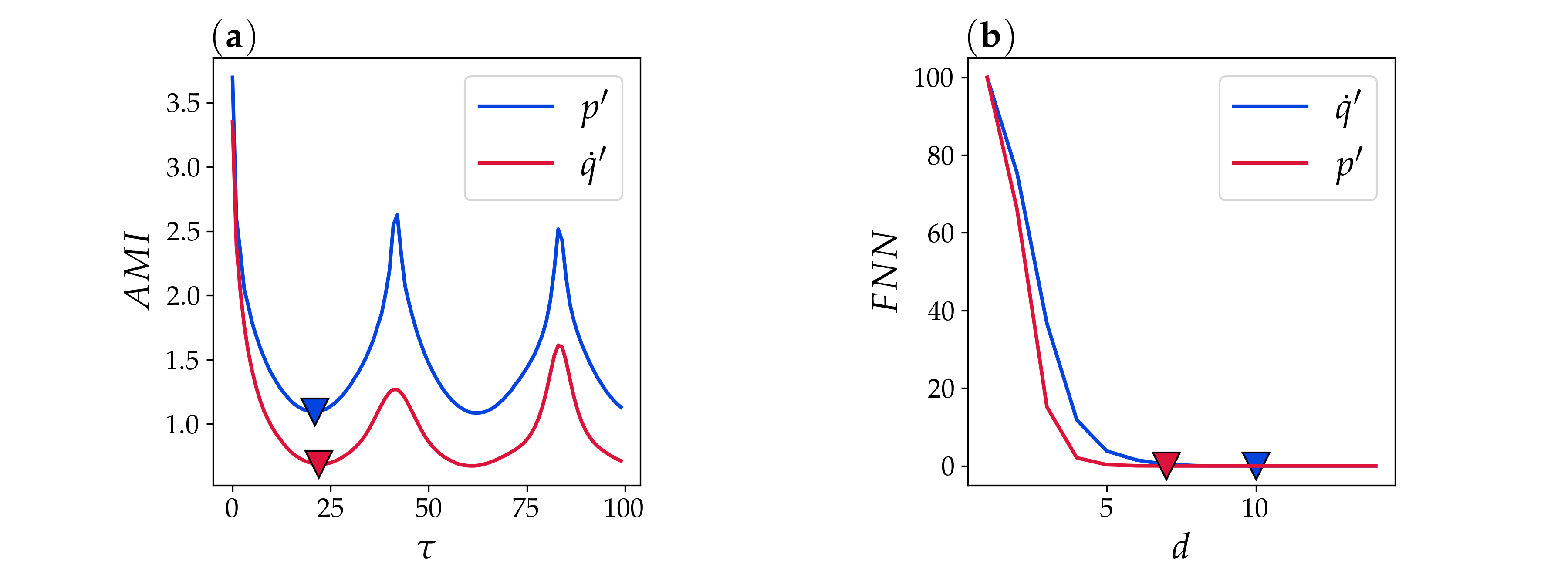}
     \caption{Parameters for the reconstruction of the phase space are obtained using average mutual information ($AMI$) and false nearest neighbor ($FNN$) method. a) Variation of $AMI$ with delay $\tau$ (number of data points). The $AMI$ attains first minima at $\tau_p$ = 21 for $p'$, $\tau_q$ = $22$ for $\dot{q}$ time series and is chosen as the optimal delay for the respective time series. Marker shows $\tau_p$ and , $\tau_q$. b) The variation of the percentage of points having $FNN$ with the dimension of phase space $d$.  The markers show the chosen dimension. The $FNN$ reduces to zero at $d_q$ = $10$ for $\dot{q}$ and at $d_p$ = $7$ for $p'$ time series. The combine dimension $d$ = $17$ is chosen as the embedding dimension for the reconstruction of the phase space.}
     \label{fig:delay_emb_TA}
\end{figure}

The AMI between the original and delay time series is computed as \cite{AMI_galler_1968},
$$
AMI(\tau) = \sum_{p(t_i),p(t_i+\tau)} P(p(t_i),p(t_i+\tau)) 
            \log_2\bigg[ \frac{P(p(t_i),p(t_i+\tau))}{P(p(t_i))P(p(t_i+\tau))}  \bigg],
$$
where, $P(p(t_i),p(t_i+\tau))$ is the joint probability while   $P(p(t_i))$ and $P(p(t_i+\tau))$ are the marginal probability of $p(t_i)$ and $P(p(t_i+\tau))$. Hence, the minimum $\tau$ for which AMI between the original and delay signal is minimum is used as the optimal delay \cite{abarbanel_1993}.

The co-ordinates of the dynamical system at time instance $t_i$ with embedding dimension $d$ is $y_d(t_i)=[p(t_i), p(t_i+\tau),.., p(t_i + (d-1)\tau)]$. We use the method of false nearest neighbors (FNN) \cite{abarbanel_1993} to determine the embedding dimension. For all the time instances, the corresponding nearest neighbor is obtained; i.e., $y_d^{(n)}(t_i)$ is nearest neighbor to $y_d(t_i)$. The Euclidean norm of the separation between $y_d(t_i)$ and $y_d^{(n)}(t_i)$ is $R_d(t_i)=|y_d(t_i) - y_d^{(n)}(t_i)|_2$. For the $d$ dimensional phase space the $y_d(t_i)$ and $y_d^{(n)}(t_i)$ are false neighbors if,
$$
\frac{[R_{d+1}^2(t_i)-R_{d}^2(t_i)]^{1/2}} {R_{d}(t_i)}> R_T
$$
For, the present analysis, $R_T$ is chosen as 10. This criterion is useful for clean data. For noisy data, if $({R_{d+1}(t_i)}/{\sigma_p}) \geq 2 $, then points are considered as false neighbors. Here, $\sigma_p$ is the standard deviation of the original time series $p(t)$.

The variation of the the $AMI$ with delay ($\tau$) and $FNN$ with embedding dimension $d$ are shown in Fig. \ref{fig:delay_emb_TA}. The first minima of $AMI$ occurs at $\tau_q$ = 22 for $\dot{q}$ and occurs at $\tau_p$ = 21 for $p'$; these values of $\tau_q$ and $\tau_p$ are chosen as optimal delay. The number of $FNN$ approaches zero for $d_q=10$ and for $d_p=7$ for time series of $p'$ and $\dot{q}$ respectively and the combined dimensions $d= d_q+d_p$ is chosen as the number of embedding dimensions to reconstruct the phase space.

\subsection{Convergence of neural ODE}{\label{subsec:conv_neural_ODE_hid}}
\begin{figure}[th!]
     \centering
         \centering
         \includegraphics[width=0.5\textwidth]{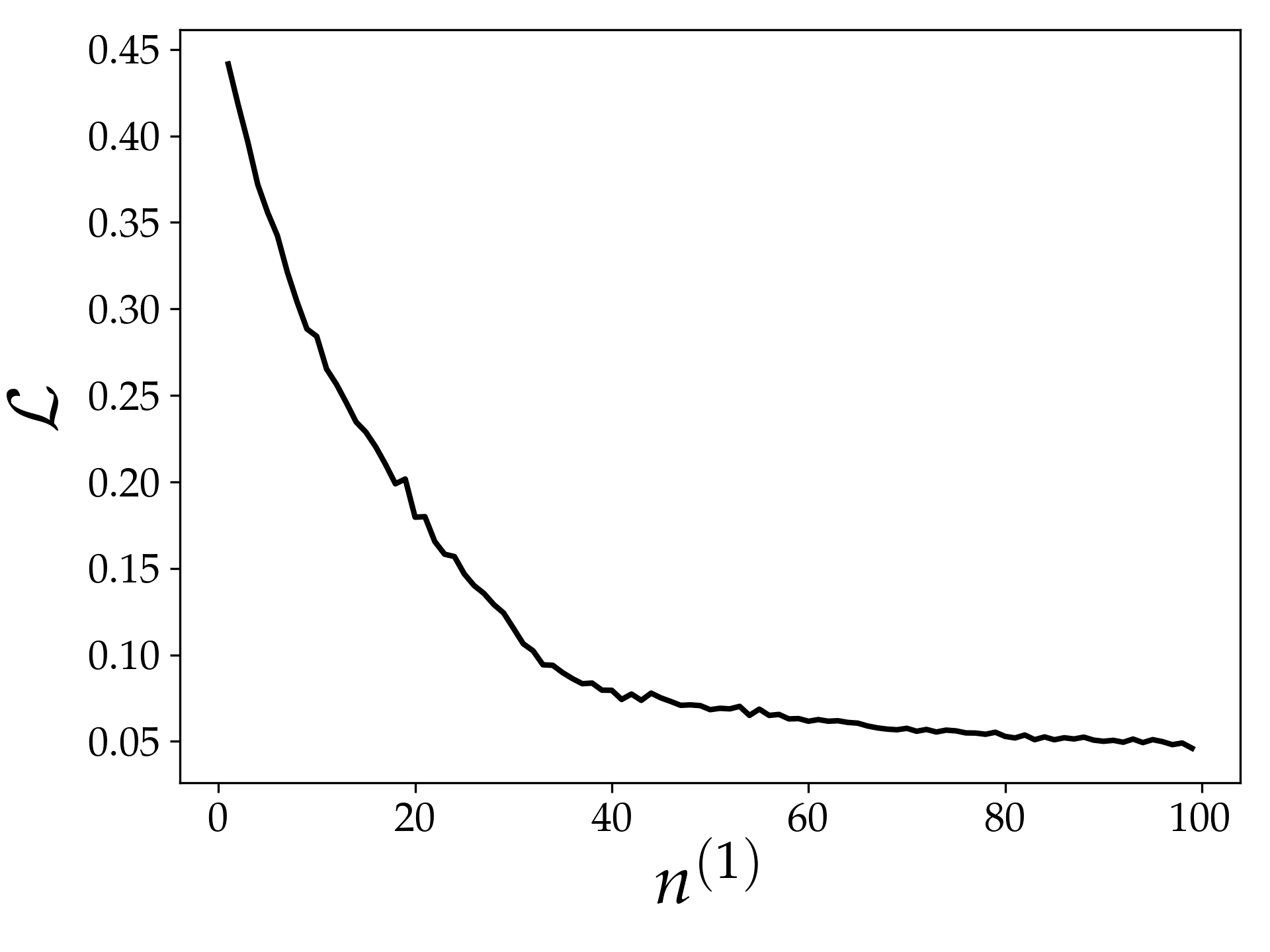}
     \caption{Variation of the average loss $\mathcal{L}$ with number of nodes in the hidden layer $n^{(1)}$. $\mathcal{L}$ is computed after training neural ODE for 2000 iterations for each $n^{(l)}$.}
     \label{fig:loss_vs_n_hid}
\end{figure}
Neural network used in the present analysis contains single hidden layer. The average loss incurred by considering different number of nodes $(n^{(l)})$ is shown in the Fig. \ref{fig:loss_vs_n_hid}. The average loss is computed after training neural ODE on time series of limit cycle oscillation with fixed 2000 number of iterations. The average loss drops rapidly with $n^{(l)}$ initially and approaches steady value. Thus, we have chosen 100 nodes in the hidden layer.

\subsection{Generalizability of the proposed framework} {\label{res_generlizability}}
\begin{figure}[ht!]
     \centering
         \includegraphics[width=0.8\textwidth]{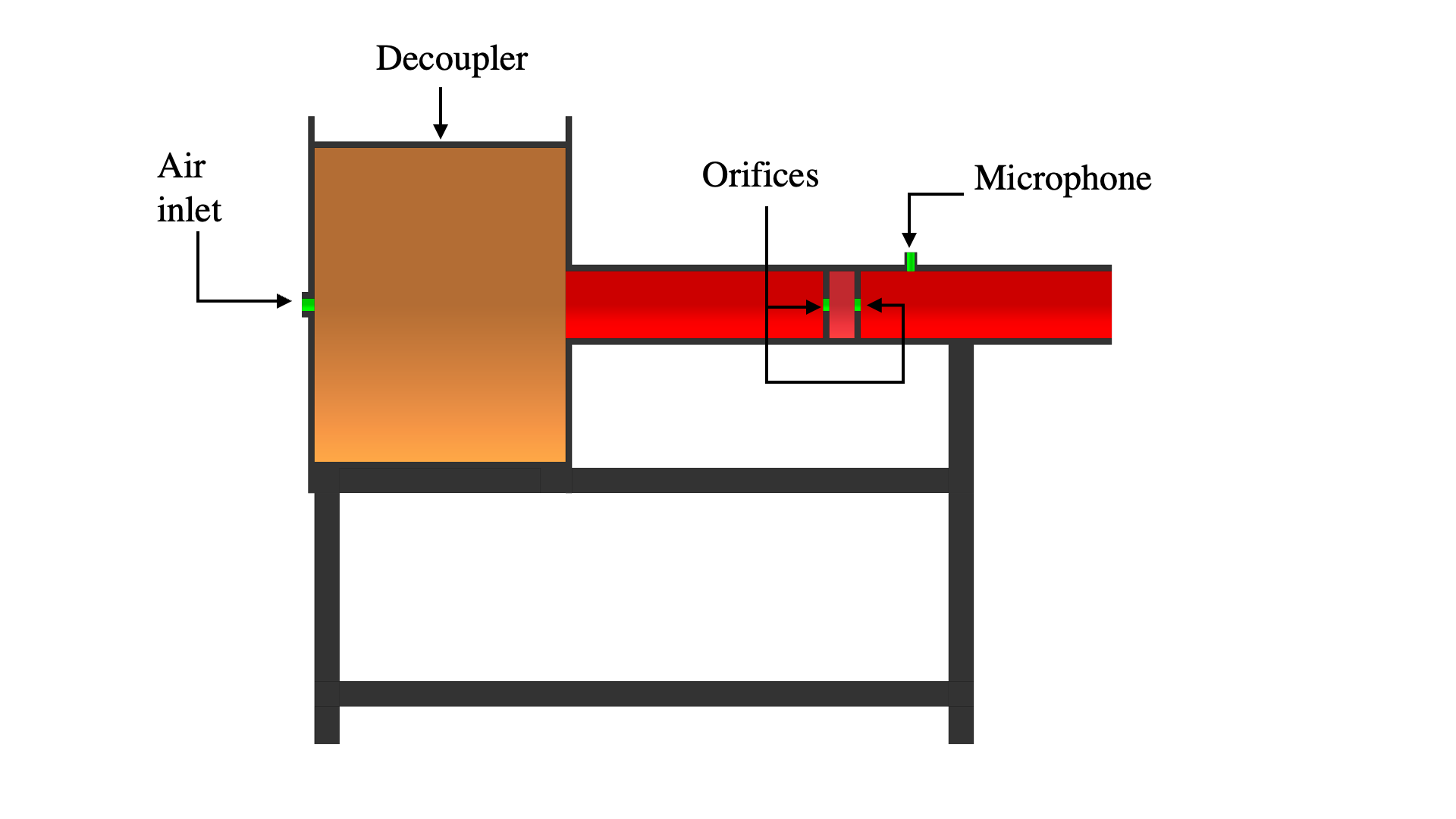}
        \caption{The aeroacoustic system consists of duct with two orifices which leads to vortex shedding. The controlled airflow rate is used to attain different dynamical states. Microphone measures the acoustic pressure fluctuations downstream of the orifice for different dynamical states.}
        \label{fig:Aeroacoustic_aeroelastic_setup}
\end{figure}

The phenomenon of transition to oscillatory instability from a state of chaotic oscillations is observed in a variety of systems. Here, we apply the proposed framework to detect the onset of oscillatory instability in aeroacoustic systems. Instabilities in aeroacoustic systems arise due to the feedback between the vortex shedding in turbulent flow and acoustic field in the confinement \citep{Flandro_2003}. Pleasant sound from wind instruments \cite{howe_1975} and oscillations established in gas pipelines \citep{KRIESELS1995343} due to feedback mechanism between shear layer, caused by the flow past the opening of the closed branch, and resonant acoustic mode \cite{Devis_closed_branches} are some examples of aeroacoustic systems. Even though the underlying mechanisms for thermoacoustic and aeroacoustic instabilities are different, we observed that the weight matrix shows similar characteristics for aeroacoustic systems as that of thermoacoustic systems.  

The experimental setup of an aeroacoustic system consists of a duct with two orifices as shown in Fig.  \ref{fig:Aeroacoustic_aeroelastic_setup}. The experimental data used in this study are reported in \citep{Pavithran_2020}. The duct involves two pipes of lengths 300 mm and 225 mm respectively and two circular orifices of diameter 20 mm separated by a distance of 18 mm. The decoupler, a large cylindrical chamber installed upstream of the duct, isolates the duct from upstream pressure fluctuations. This large chamber maintains the ambient pressure at both ends of the duct. The mass flow controller (Alicat MCR series) is used to regulate the air mass flow rate with an uncertainty of $\pm$($0.8\%$ of reading + $0.2\%$ of full scale). During the experiment, the mass flow rate of air is varied from $1.633\pm 0.054 $ g/s to $2.695 \pm 0.062$ g/s in step of $0.041$ g/s. The corresponding $Re$ varies from 5615 $\pm$ 185 to 9270 $\pm$ 212. The pressure measurements within the duct are performed using a pre-polarized microphone and preamplifier system (PCB make, model number 378C10, 1 mV/Pa sensitivity and 28.3 $\mu$Pa resolution) with a sampling frequency of 10kHz. The microphone is installed at a distance of 100 mm from the second orifice. 

\begin{figure}[ht]
     \centering
         \centering
         \includegraphics[width=0.85\textwidth]{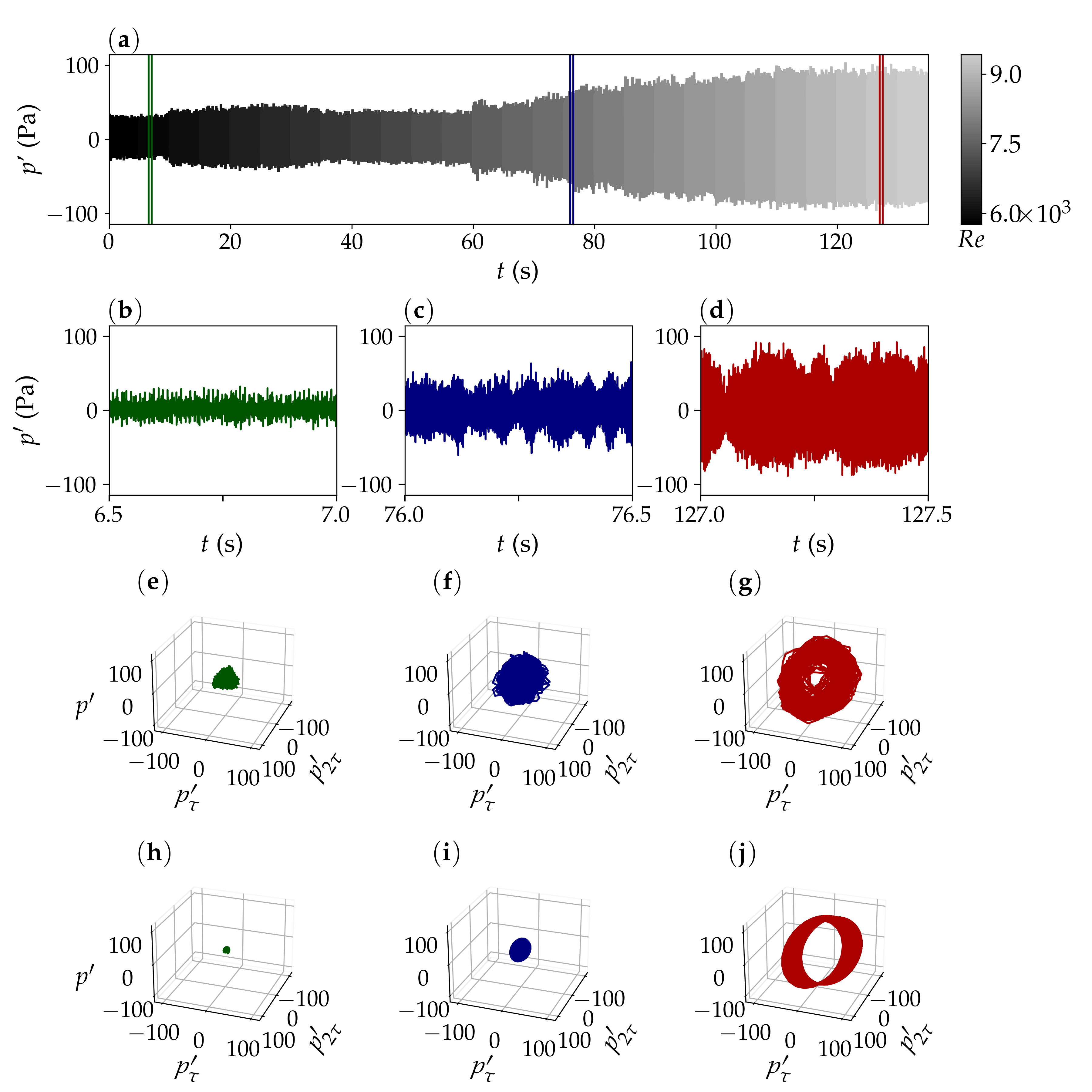}
     \caption{Time series of state variable for transition from chaotic to oscillatory instability for aeroacoustic system as shown in Fig. \ref{fig:Aeroacoustic_aeroelastic_setup} (a). The sampling frequency is 10kHz and for each operating state 5 s data is acquired. (a) The 5 s data acquired for each operating state is stacked to make continuous time series representing the transition. (b), (c), (d) shows 0.02 s time series at $Re$ = $6326$, $7451$, $9274$ respectively. (e), (g), (i) shows the phase portrait of time series shown in (b), (c), (d) respectively in delay coordinate $( p'(t),{p}'(t+\tau), p'(t+ 2\tau))$ coordinates. (f), (h), (j) shows the trajectory in phase space obtained using neural ODE with the the same initial condition as that of phase trajectory in (e), (g), (i) respectively.}
     \label{fig:res_phase_aeroacoustic}
\end{figure}

\begin{figure}
     \centering
        \includegraphics[width=\textwidth]{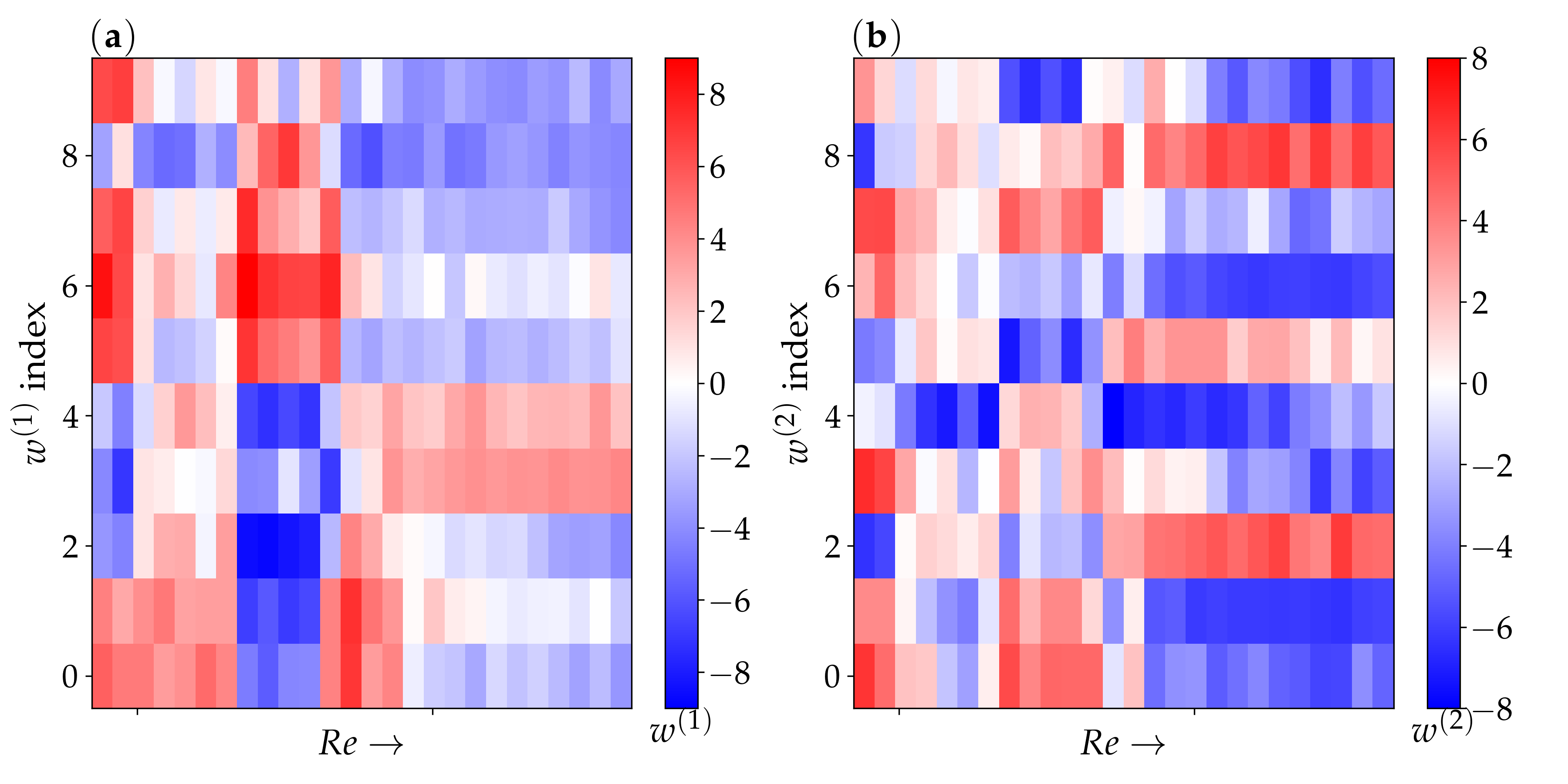}
     \caption{Evolution of the ten elements of weight matrix with Reynolds number for aeroacoustic system. These ten elements are chosen from each weight matrix with highest variance. (a) Evolution of ten elements from $W^{(1)}$ matrix, (b) Evolution of ten elements of $W^{(2)}$, with highest variance. Color bar indicates the value of weight matrix elements.} 
     \label{fig:res_precursor_aeroacoustic_color}
\end{figure} 

\begin{figure}
     \centering
        \includegraphics[width=\textwidth]{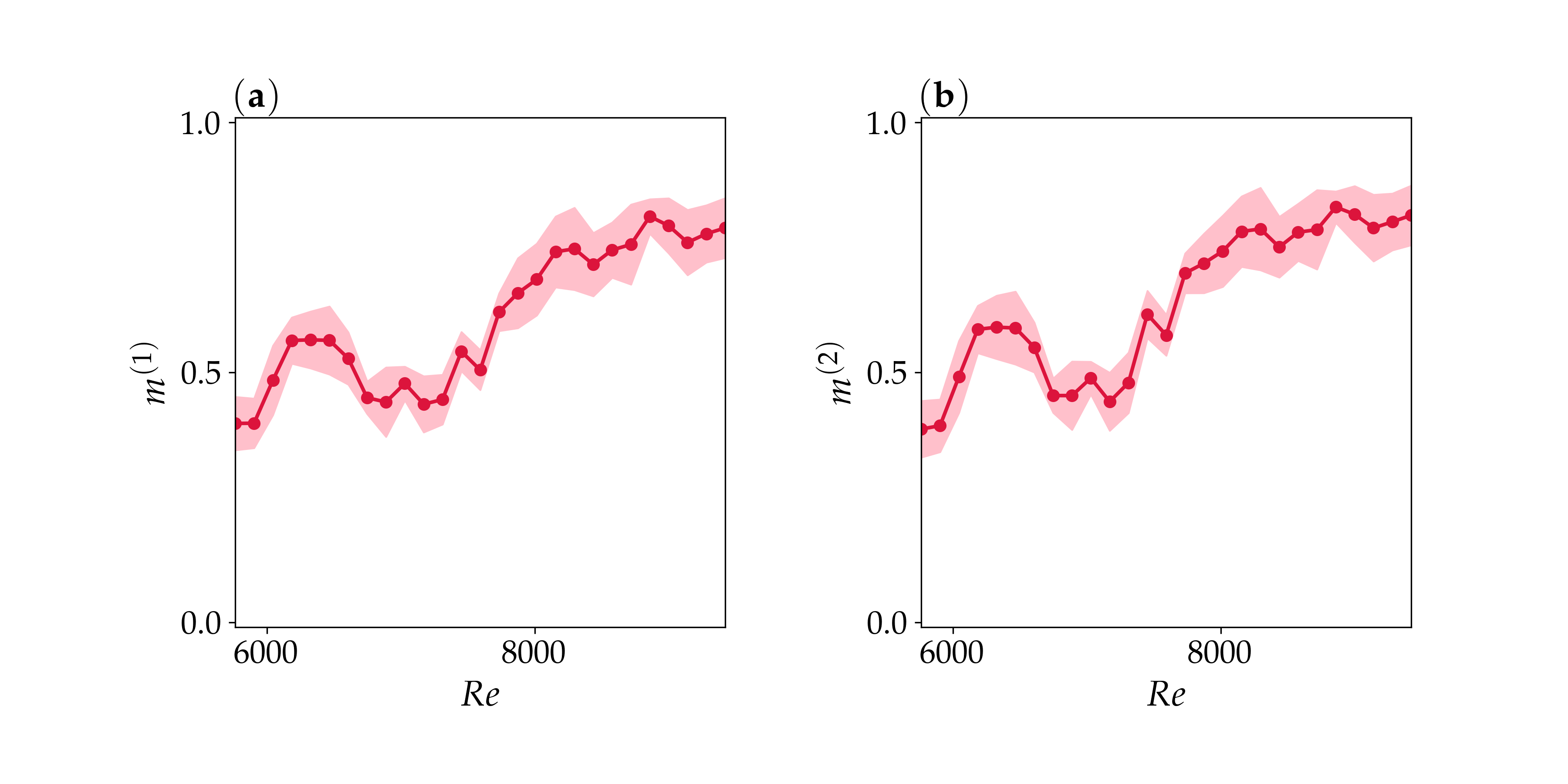}
     \caption{Variation of the measure $m^{(l)}$ with Reynolds number for $l^{th}$ layer of neural ODE for aeroacoustic system. (a) Variation of $m$ for the first weight matrix ($W^{(1)}$), (b) Variation of $m$ for  second weight matrix ($W^{(2)}$). The meaning of all the terms is the same as that described in Fig. \ref{fig:res_precursor}} 
     \label{fig:res_precursor_aeroacoustic}
\end{figure} 

The time series of pressure fluctuations acquired from the aeroacoustic system is shown in Fig. \ref{fig:res_phase_aeroacoustic} (a). The time series shows the transition from chaos to oscillatory instability on changing the $Re$. Enlarged views of the time series for chaos, intermittency and limit cycle oscillations are shown in the figures (b), (c) and (d) of Fig. \ref{fig:res_phase_aeroacoustic}. For the reconstruction of the phase space, $\tau=$ 5 and embedding dimension $d =$ 8 are obtained by the method discussed in Sec. \ref{subsec:model_HR}. Figures \ref{fig:res_phase_aeroacoustic} (e), (f) and (g) show the reconstructed phase portrait for the time series data shown in figures (b), (c) and (d) respectively. Training of neural ODE on data for a sequence of $Re$ gives the neural ODE parameter vector $\theta(Re)$. Figures \ref{fig:res_phase_aeroacoustic} (h), (i), (j) show the predicted phase trajectory for the thermoacoustic system with the initial condition same as that of the trajectories shown in figures \ref{fig:res_phase_aeroacoustic} (e), (f), (g). The neural ODE is integrated with the ${\theta}$ obtained for the respective operating states. Comparing the phase portrait with their corresponding phase portrait reconstructed from experimental data, we see that the neural ODE can reconstruct the phase portrait for the aeroacoustic system.   

Similar to the thermoacoustic system, the evolution of the parameter vector $\theta$ is analyzed. Figure \ref{fig:res_precursor_aeroacoustic_color} shows the variation of the first ten elements of weight matrices having the highest variance with $Re$. The variation of the elements in $W^{(1)}$ and $W^{(2)}$ are shown in figures \ref{fig:res_precursor_aeroacoustic_color} (a) and (b) respectively. The elements of the weight matrix $W^{(1)}$ show a discernible pattern that most of the weights have a high magnitude for the state of chaos and it approaches zero as the aeroacoustic system approaches the state of limit cycle oscillations. In contrast, elements of the weight matrix $W^{(2)}$ show the opposite behavior; i.e., most of the weight elements have smaller magnitude for the state of chaos and their magnitude increases as the thermoacoustic system approach limit cycle oscillations. A similar feature is observed in Fig. \ref{fig:res_ta_color_map}.          

The variation of the precursor measure $m^{(l)}$ for the layer $(l)$ is shown in the Fig. \ref{fig:res_precursor_aeroacoustic} (a) and (b) respectively. The cosine measure $m^{(l)}$ approaches a value of one as the aeroacoustic system approaches the state of periodic oscillations. 

Thus, we propose that the proximity of the thermoacoustic or aeroacoustic system, exhibiting a transition from chaos to periodic oscillations, can be predicted using ${m}^{(l)}$. The position of the thermoacoustic or aeroacoustic system in the parametric space, defined with $\theta$ as coordinates, changes with the change in the system dynamics. $m^{(l)}$ approaches one as the system approaches the state of limit cycle oscillations.

\section*{Data Availability Statement}

Data will be made available on reasonable request.

\nocite{*}
\bibliography{aipsamp}

\end{document}